\documentclass[onecolumn,aps,pre]{revtex4}
\usepackage{graphicx}
\usepackage[]{epsfig}
\usepackage{amsmath,amsthm,amssymb,amsthm,mathrsfs}
\usepackage{hyperref}
\usepackage{paralist}
\usepackage{bm}

\def\<{\langle}
\def\>{\rangle}
\def\(({\left(}
\def\)){\right)}
\def\[[{\left[}
\def\]]{\right]}

\newcommand{\beq}{\begin{equation}}
\newcommand{\eeq}{\end{equation}}
\newcommand{\beqd}{\begin{displaymath}}
\newcommand{\eeqd}{\end{displaymath}}
\newcommand{\beqa}{\begin{eqnarray}}
\newcommand{\eeqa}{\end{eqnarray}}

\newcommand{\comment}[1]{}

\begin{document}

\title{Comment on `Real-space renormalization-group methods for hierarchical spin glasses'}

\author{Maria Chiara Angelini $^{1}$, Giorgio Parisi $^{1,2,3}$, Federico Ricci-Tersenghi $^{1,2,3}$}

\affiliation{$^1$Dipartimento di Fisica, Sapienza Universit\`a di Roma, Piazzale A. Moro 2, I-00185, Rome, Italy}
\affiliation{$^2$ Nanotec-CNR, UOS Rome, Sapienza Universit\`a di Roma, Piazzale A. Moro 2, I-00185, Rome, Italy}
\affiliation{$^3$ INFN-Sezione di Roma 1, Piazzale A. Moro 2, 00185, Rome}

\begin{abstract}
In the paper [Angelini M C, Parisi G, and Ricci-Tersenghi F, \textit{Ensemble renormalization group for disordered
systems}, Phys. Rev. B \textbf{87} 134201 (2013)] we introduced a real-space renormalization group called Ensemble Renormalization Group (ERG) and we applied it to the Edwards-Anderson model, obtaining estimates for the critical exponents in good agreement with those from Monte Carlo simulations.
Recently the paper [Castellana M, \textit{Real-space renormalization-group methods for hierarchical spin glasses}, J. Phys. A: Math. Theor. \textbf{52} 445002 (2019)] re-examined the ERG method from a different perspective, concluding that the previous results were wrong, and claiming that the ERG method predicts trivially wrong critical exponents.
In this comment we explain why the conclusions reached by Castellana are wrong, as they are based on a misinterpretation of finite-size effects. We conclude that the ERG method remains a good RG method to obtain critical exponents in strongly disordered models (if properly used).
\end{abstract}

\maketitle

The renormalization group (RG) has been a wildly used tool to study and characterize phase transitions. 
Unfortunately, its application to disordered systems has encountered many difficulties. 
Among the disordered systems, a particularly challenging model is the Edwards-Anderson (EA) spin-glass model \cite{EA}. 
In Ref.~\cite{ERG} a real-space RG method, called Ensemble RG (ERG), has been introduced for the EA model on the Dyson lattice \cite{Dyson}, i.e.\ a particular one-dimensional lattice with long-range interactions whose strength decays as an inverse power of distance. Changing the power-law exponent, the hierarchical lattice can mimic a regular $d$-dimensional lattice in its long distance critical properties.
When applied to the EA model on the Dyson lattice, the ERG method finds values for the critical exponent $\nu$, that determines the growth of the correlation length as a function of temperature $\xi\propto |T-T_c|^{-\nu}$, in accordance with theoretical and numerical estimates both above and below the upper critical dimension $d_{\rm U}=6$, as shown in Fig.~7 of Ref.~\cite{ERG}.

In a very recent paper \cite{Michele} the author examines the fixed-point structure of the ERG method for a system of finite size, concluding that the results reported in Ref.~\cite{ERG} are wrong, and claiming that the ERG method predicts an exponent $\nu$ satisfying the relation $2^{1/\nu} = 1$, that disagrees with previous theoretical and numerical estimates of $\nu$
both above and below $d_{\rm U}$. We show below that the conclusions reached in Ref.~\cite{Michele} are wrong because they are affected by the finite size effects, which have not been taken into account properly.

The EA model on the Dyson lattice, that we will call in the following hierarchical EA (HEA) model, is a $\pm1$ spin model whose Hamiltonian for $2^{k+1}$ spins can be constructed iteratively as
\beq\label{eq_H}
H_{k+1}[S_1, \cdots S_{2^{k+1}}] = H_k[S_1, \cdots, S_{2^k}] + H_k[S_{2^k+1}, \cdots, S_{2^{k+1}}] - 2^{-\zeta (k+1)} \sum_{i<j=1}^{2^{k+1}} J_{ij} S_i S_j\,.
\eeq
The couplings $J_{ij}$ at the $k+1$-th hierarchical level are independent, identically distributed random variables extracted from a Gaussian distribution with zero mean and standard deviation $\sigma_{k+1}$, that we call $P_{k+1}(J)$.
Changing the exponent $\zeta$ one can change the \textit{effective dimension} of the model.

The ERG method introduced in Ref.~\cite{ERG} works as follows:
\begin{compactenum}
\item It starts from an ensemble of systems with $n$ levels, willing to reduce this ensemble to an ``equivalent'' ensemble of smaller systems with $n-1$ levels.
\item It computes $n-1$ observables $\overline{\langle O_j \rangle}$, $j\in \{2,\ldots,n\}$ averaged over the ensemble of larger systems where couplings are distributed according to $\{P_k\}_{k\le n}$. In the HEA case, the observables are normalized spin-glass correlations.
\item \label{step2}It identifies new distributions of couplings
$P'_k(J')$, $k\in\{1,\ldots,n-1\}$ for the ensemble of smaller
systems, i.e., it determines the new $n-1$ variances of the Gaussian
distributions $P'$, by requiring that $\overline{\langle O'_i
\rangle}_{\bm{P'}}=\overline{\langle O_{i+1} \rangle}_{\bm{P}}$
for any $i \in \{1,2,...,n-1\}$.
\item \label{step3} Finally it builds a new ensemble of systems of the original size.
They are constructed joining with random couplings extracted from
the original distribution $P_n(J)$ two smaller systems with
couplings extracted from $P'_k(J')$, $k\in\{1,2,...,(n-1)\}$ found
at step \ref{step2}.
\end{compactenum}
Primed quantities refer to the smaller systems. 
Let us emphasize, as already pointed out in \cite{ERG}, that the first three steps are the true
renormalization steps, while the latter is required to obtain a final
system size, that will allow one to iterate the method, until
convergence. If one could deal with an infinite size system, there will be no need for step \ref{step3}: this subtlety is responsible for the wrong conclusions reached in Ref.~\cite{Michele}.
In Ref.~\cite{ERG}, thermal averages are computed exactly by
exhaustive enumeration, thus limiting the analysis to a small number of levels $n=4$. 

In Ref.~\cite{Michele} the author computes the matrix corresponding to the linearized RG equation
\begin{eqnarray}
\label{M}
M= \left(\begin{array}{cccc}
\frac{\partial \sigma^{t+1}_1}{\partial \sigma^t_1} & \multicolumn{2}{c}{\cdots} & \frac{\partial \sigma^{t+1}_1}{\partial \sigma^t_n}\\
\vdots & && \vdots\\
\frac{\partial \sigma^{t+1}_{n-1}}{\partial \sigma^t_1} & \multicolumn{2}{c}{\cdots} & \frac{\partial \sigma^{t+1}_{n-1}}{\partial \sigma^t_n}\\
0 & \cdots & 0 & 1\\
\end{array}\right),
\end{eqnarray}
where $\sigma_i^t$ indicates the variance of the Gaussian distribution of the couplings at level $i$ after $t$ steps of ERG and the last row of the matrix is due to the fact that $\sigma_n^{t+1}=\sigma_n^{t}=\sigma_n^0$, as imposed by step \ref{step3} of ERG.
At this point one can compute the eigenvalues of $M$ obtaining $\lambda_1=1$ due to the bottom-right $1\times 1$ block, plus $\lambda_2, \cdots, \lambda_n$ 
from the top-left, $(n-1) \times (n-1)$ block of $M$. In \cite{Michele}, the eigenvalues are numerically evaluated in the case $n=4$, for the whole interesting range of the coupling exponent $\zeta$,
always obtaining $|\lambda_i|<1$, for $i=2,3,4$. At this point, using a known RG relation introduced in \cite{Wilson}, the author of \cite{Michele} reaches the (wrong) conclusion that the ERG 
prediction is $2^{1/\nu}=\lambda_1=1$. This prediction disagrees with other estimates of $\nu$ in the whole range of $\zeta$ (and thus for all effective dimensions $d$), and the conclusion of \cite{Michele}
seems to be that the ERG method is not a good RG method.
However, the relation that links the exponent $\nu$ to the eigenvalue with the largest norm $\lambda_{max}$ of the linearized matrix $M$, $2^{1/\nu}=\lambda_{max}$, 
cannot be applied to the matrix $M$ in eq. \ref{M} as done in \cite{Michele}. This is because in this case the eigenvalue with the largest norm is always $\lambda_{max}=\lambda_1=1$, the presence of
this eigenvalue $\lambda_1$ is due to step \ref{step3}
of the ERG method. However, as we already stressed, step \ref{step3} is just introduced to iterate the procedure, given the finite size of the analyzed systems, and it is thus the step responsible for finite-size effects in the ERG results. The prediction $2^{1/\nu}=1$ in \cite{Michele} is thus completely dominated by finite-size effects.

In Ref.~\cite{ERG}, a different method was used to extract $\nu$, looking at how fast the distance between two RG trajectories at different inverse temperatures $\beta_1$ and $\beta_2$ grows with the RG step $t$.
Following \cite{Wilson},
\beq
\frac{\beta_1 \sigma^{t}_1(\beta_1) - \beta_2 \sigma^{t}_1(\beta_2)}{\beta_1 - \beta_2} \propto 2^{t/\nu}\,,
\label{eq:correct_nu}
\eeq
if $\beta_1$ and $\beta_2$ are in the vicinity of the critical fixed point (FP).
Looking at the renormalized flow of $\sigma^t$, shown in Fig.~5 of Ref.~\cite{ERG} or in Fig.~1 of Ref.~\cite{Michele}, one can notice --- as already pointed out in \cite{ERG} --- that, starting above the critical temperature, $\sigma_t$ decays and then, after a small number of steps $t^*$, reaches a ``fixed point'' that is not the usual $\sigma=0$ high-temperature FP, but has $\sigma_{FP}>0$ and depends on the starting temperature.
Analogously, starting below the critical temperature, $\sigma_t$ grows and after $t^*$ steps reaches a ``fixed point'' that is not the usual $\sigma=\infty$ low-temperature FP, but depends on the starting temperature. 
The reason for the existence
of this plateau after $t^*$ is again the step \ref{step3} of the ERG procedure and it is thus completely unphysical, 
but it is just the consequence of an approximation introduced because of the finite-size of the analyzed systems.
If we were dealing with systems of infinite size, starting below the critical temperature, $\sigma^t$ would continue to grow indefinitely for $t\to\infty$ and the plateau would not exist.
For this reason in Ref.~\cite{ERG}, eq.~(\ref{eq:correct_nu}) was used to extract $\nu$, fitting data only for $t<t^*$, avoiding in this way to fit the ``unphysical'' part of the RG flow affected by finite size effects.
In the same paper it was also shown that $t^*$ grows if larger systems are used (for the ferromagnetic version of the model, the ERG method was applied up to $n=13$), going to $t^*=\infty$ in the $n\to\infty$ limit. In this limit,
step \ref{step3} is no more needed in the ERG procedure, and finite size effects disappear. 

The method for computing $\nu$ of Ref.~\cite{Michele} corresponds to the use of eq.~(\ref{eq:correct_nu}) in the limit $t\to\infty$, as the author correctly pointed out. But taking this limit on systems of finite size, corresponds to study the evolution of the RG flow on the ``unphysical'' plateau dominated by finite size effects, thus leading to an unphysical result for $\nu$. 
In \cite{Michele}, the author correctly pointed out that fitting data with eq.~(\ref{eq:correct_nu}) for $t<t^*$, as we did in Ref.~\cite{ERG}, takes contribution not only from $\lambda_1$ but also from $\lambda_i$ with $i=2,3,4$ for $n=4$. This is exactly the correct thing to do, because, in the large $n$ limit, where step \ref{step3} of ERG is no more needed, the 
matrix $M$ will not have the $\lambda_1=1$ eigenvalue, but only the others eigenvalues $\lambda_i$ with $2\le i \le n$.
In Ref.~\cite{ERG} we showed how well an exponential fit of the form (\ref{eq:correct_nu}) fits the data obtained for the ferromagnetic model with $n=13$ in a much broader regime $0<t\le 35 <t^*$, giving a value for $\nu$ that is in perfect agreement
with the exact result.

Concluding, the ERG method proposed in \cite{ERG} to compute critical exponents is correct, because it considers only the ``physical'' part of the renormalization flow that is not dominated by finite size effects. Conversely, the results of \cite{Michele} are wrong, because completely dominated by the approximations made to run ERG on systems of finite size.
The ERG method, as used in \cite{ERG}, is a valid method to obtain critical properties in strongly disordered systems.

This research has been supported by the European Research Council under the European Unions Horizon2020 research and innovation programme (grant No. 694925 – Lot-glassy, G Parisi)

\end{document}